\documentclass[iop]{emulateapj}

\newcommand{\hII}{H\,{\sc ii} }
\newcommand{\pv}{$p-v$ }
\slugcomment{}

\shorttitle{Sh 2-53 region: accreting molecular filaments}
\shortauthors{T. Baug et al.}
\begin{document}

\title{Star formation in the Sh 2-53 region influenced by accreting molecular filaments}
\author{T. Baug}
\affil{Aryabhatta Research Institute of Observational Sciences (ARIES), Manora Peak, Nainital 263001, India\\
Department of Astronomy and Astrophysics, Tata Institute of Fundamental Research, Homi Bhabha Road, Mumbai-400005, India}

\author{L. K. Dewangan}
\affil{Physical Research Laboratory, Navrangpura, Ahmedabad 380009, India.}

\author{D. K. Ojha}
\affil{Department of Astronomy and Astrophysics, Tata Institute of Fundamental Research, Homi Bhabha Road, Mumbai 400005, India}

\author{Kengo Tachihara}
\affil{Department of Physics, Nagoya University, Chikusa-ku, Nagoya 464-8602, Japan}

\author{A. K. Pandey}
\affil{Aryabhatta Research Institute of Observational Sciences (ARIES), Manora Peak, Nainital 263001, India}

\author{Saurabh Sharma}
\affil{Aryabhatta Research Institute of Observational Sciences (ARIES), Manora Peak, Nainital 263001, India}

\author{M. Tamura}
\affil{Department of Astronomy, The University of Tokyo, 7-3-1, Hongo, Bunkyo-ku, Tokyo 113-0033, Japan \\
Astrobiology Center of NINS, 2-21-1, Osawa, Mitaka, Tokyo 181-8588, Japan \\
National Astronomical Observatory of Japan, 2-21-1, Osawa, Mitaka, Tokyo 181-8588, Japan}

\author{J. P. Ninan}
\affil{Department of Astronomy and Astrophysics, The Pennsylvania State University, University Park, USA}

\and

\author{S. K. Ghosh}
\affil{Department of Astronomy and Astrophysics, Tata Institute of Fundamental Research, Homi Bhabha Road, Mumbai 400005, India\\
 National Centre for Radio Astrophysics, Tata Institute of Fundamental Research, Pune 411007, India}

\email{tapas.polo@gmail.com}

\begin{abstract}
We present a multi-wavelength analysis of a $\sim$30$' \times$30$'$ area around the Sh 2-53 region (hereafter S53 complex),
 which is associated with at least three \hII regions, two mid-infrared bubbles (N21 and N22), and infrared dark clouds.
 The $^{13}$CO line data trace the molecular content of the S53 complex in a velocity range of 36--60 km s$^{-1}$,
 and show the presence of at least three molecular components within the selected area along this direction. Using the
 observed radio continuum flux of the \hII regions, the derived spectral types of the ionizing sources agree well with the
 previously reported results. The S53 complex harbors clusters of young stellar objects (YSOs) that are identified using the
 photometric 2--24 $\mu$m magnitudes. It also hosts several massive condensations (3000-30000 $M_\odot$) which are traced in
 the {\it Herschel} column density map. The complex is found at the junction of at least five molecular filaments, and the
 flow of gas toward the junction is evident in the velocity space of the $^{13}$CO data. Together, the S53 complex is embedded
 in a very similar ``hub-filament" system to those reported in \citet{myers09}, and the active star formation is evident towards
 the central ``hub'' inferred by the presence of the clustering of YSOs.
\end{abstract}
\section{Introduction}
\label{sec:intro}
Formation mechanism of massive stars ($>$8 M$_\odot$) is an unresolved problem in astrophysics in spite of the fact that they
 have significant influence to determine the fate of their host galaxies through strong stellar wind, energetic ultra-violet (UV)
 radiation and supernova explosion. Several theoretical models have been proposed for describing the formation of massive stars
 \citep[see][for more details]{motte17,tan14}, but none of these theories are yet very well accepted. After the availability of the
 {\it Herschel} observations, filamentary structures are frequently identified towards the star-forming regions, and
 they are often considered to play an important role in the star formation processes \citep{schneider12, dale11,myers09}. There
 are increasing evidences for these filaments to channel the molecular gas toward the central junction, where the active star
 formation (including the formation of massive stars) is occurring \citep[see][and references therein]{dewangan17,baug15,nakamura14}.
 It is also often found that filaments themselves harbor \hII regions and methanol masers, which are obvious signatures of
 massive star formation \citep[see][]{schneider12,dewangan16a,dewangan17}.
 
It is also known that the energetics from massive Wolf-Rayet and O-stars have the ability to influence the surrounding molecular
 environment to form next generation stars \citep[e.g.,][]{dewangan16b,samal10}. However, it is not yet well established 
 observationally how exactly a massive star triggers the formation of new generation of stars. A detailed discussion about the various
 processes of triggered star formation can be found in the review article by \citet{elmegreen98}. Detailed observational studies
 are also performed showing the influence of massive stars on their surroundings inferring triggered star formation, but only for
 a handful of regions \citep[see][and references therein]{samal10, zavagno10a, zavagno10b, pomares09}.
 Hence, in parallel to study the formation mechanism, understanding the influence of massive stars on their surroundings is also
 equally important. If not identified directly, presence of a massive star can always be inferred by the existence of \hII regions
 or 6.7 GHz Class II methanol masers in a given star-forming cloud. They are also often linked with the Galactic mid-infrared (MIR)
 bubbles, which have been widely identified in the {\it Spitzer} 8 $\mu$m images \citep{churchwell06,churchwell07,simpson12}. 

In this paper, we have selected a $\sim$30$'\times$30$'$ area centered on $l=$ 18$^\circ$.140, $b=$-0$^\circ$.300
 toward the Sh 2-53 region to probe the ongoing star formation scenario. The selected region is associated with at least
 three \hII regions, two MIR Galactic bubbles \citep[N21 and N22;][] {churchwell06}, and infrared dark clouds (hereafter
 S53 complex). The paper is organized in the following manner. We describe the multi-wavelength data and their analyses
 in Section~\ref{sec:observations}. More details on the past studies, morphology, and the open questions on this region
 are described in section~\ref{sec:morphology}. The main results of this study are presented in Section~\ref{sec:surrclouds}.
 A detailed discussion on the possible star formation scenario operating in the S53 complex and its surrounding region
 is elaborated in Section~\ref{sec:SF}, and a conclusion of the study is presented in Section~\ref{sec:conclusions}.

\section{Observations and data reduction}
\label{sec:observations}
To understand the ongoing star formation processes in the S53 complex, we have utilized the available multi-wavelength data
 starting from the near-infrared (NIR) to radio frequency. The details of these multi-wavelength data are briefly described in
 the following sub-sections.

\subsection{Near-infrared Imaging Data}
In order to identify and classify young stellar objects (YSOs) toward the S53 complex, the NIR photometric
 magnitudes of point-like sources are obtained from the 3.8-m United Kingdom Infrared Telescope (UKIRT) Infrared Deep Sky
 Survey (UKIDSS) Galactic Plane Survey archive \citep[GPS release 6.0; ][]{lawrence07}. The UKIDSS images have a spatial
 resolution of $\sim$0$\farcs$8. The sources with good photometric magnitudes, having accuracy better than 10\%, are only
 considered for the analysis of YSOs. The reliable sources are selected for the complex using the SQL criteria given in
 \citet{lucas08} and \citet{dewangan15}. In general, bright sources are saturated in the UKIDSS frames, and thus in the
 final catalog, magnitudes for the sources brighter than J = 13.25 mag, H = 12.75 mag and K = 12.0 mag are replaced by the
 Two Micron All Sky Survey \citep[2MASS;][]{skrutskie06} magnitudes \citep{cutri03}.
\subsection{Mid-infrared Data}
The MIR images, and photometric magnitudes of point-like sources are obtained from the {\it Spitzer}-Galactic Legacy Infrared
 Mid-Plane Survey Extraordinaire \citep[GLIMPSE;][]{benjamin03} survey archive. The {\it Spitzer}-IRAC images have a spatial resolution of
 $\sim$2$\arcsec$. The photometric magnitudes of point sources are obtained from the GLIMPSE-I Spring '07 highly reliable catalog.
 The Multiband Infrared Photometer for {\it Spitzer} (MIPS) Inner Galactic Plane Survey \citep[MIPSGAL;][]{carey05} 24 $\mu$m photometric
 magnitudes of point sources \citep{gutermuth15} are also used in the analysis.
\subsection{Far-infrared and millimeter data}
The level2\_5 processed {\it Herschel} 70--500 $\mu m$ images are utilized mainly to construct column density and temperature maps of the region.
 The {\it Herschel} images have beam sizes of 5$\farcs$8, 12$\arcsec$, 18$\arcsec$, 25$\arcsec$, and 37$\arcsec$ at 70, 160, 250, 350, and 500
 $\mu m$, respectively \citep{poglitsch10,griffin10}.
\subsection{Molecular line data}
In order to investigate the molecular gas related to the S53 complex in detail, we retrieve the $^{13}$CO (J=1--0) line
 data from the Galactic Ring Survey \citep[GRS;][]{jackson06}. The GRS line data have a velocity resolution of 0.21~km\,s$^{-1}$,
 an angular resolution of 45$\arcsec$ with 22$\arcsec$ sampling, a main beam efficiency ($\eta_{\rm mb}$) of $\sim$0.48, a velocity
 coverage of $-$5 to 135~km~s$^{-1}$, and a typical rms sensitivity (1$\sigma$) of $\approx0.13$~K \citep{jackson06}. 
\subsection{Radio continuum data}
We obtained the archival 610 and 1280 MHz radio continuum data from the Giant Metrewave Radio Telescope (GMRT) archive. The 610 and 1280 MHz
 data were observed on 2005 April 08 and 2005 March 12, respectively (Project Code: 07PKS01). The GMRT data have been reduced using the
 Astronomical Image Processing Software (AIPS), following the similar procedures described in \citet{mallick13}. The synthesized beam sizes
 of these 610 and 1280 MHz maps are 6$\farcs$7$\times$5$\farcs$7 and 7$\farcs$8$\times$3$\farcs$0, and rms of 1.0 mJy/beam and 0.5 mJy/beam,
 respectively.
\section{The selected region and morphology}
\label{sec:morphology}
The Sh 2-53 region is located at a distance of 4.0 kpc \citep{paron13}. Using $^{13}$CO data, \citet{paron13} reported
 the existence of a large molecular shell \citep[$\sim$70 pc $\times$ 28 pc; see Figure 9 of][]{paron13}, which was traced in
 a velocity range from 51--55 km s$^{-1}$, and they have pointed out that the S53 complex is situated near the edge of the
 large molecular shell. Also a supernova remnant \citep[SNR G18.1--0.1;][]{green14} is seen close to the center of the large
 molecular shell. However, \citet{paron13} discarded any connection of the SNR with the large shell. Later, \citet{leahy14, kilpatrick16}
 found that the SNR is located at a different distance (5.6 kpc) with a higher local standard rest velocity (V$_{LSR} \sim$ 100 km s$^{-1}$).
 A few small scale studies are available mainly to address the star formation scenario of the individual \hII regions separately
 \citep[see e.g.,][]{watson08, ji12, sherman12}. However, a detailed multi-wavelength analysis of the large area to address
 the overall star formation scenario is not yet explored. For example, the analysis of the dust clumps using the {\it Herschel}
 data is not performed. Moreover, the dynamics of the molecular gas is yet to be analyzed carefully. The complex is also reported
 to be associated with several photometrically identified massive stars. However, none of the previous studies addressed the
 formation mechanism of massive stars and the S53 complex.

\begin{figure*} 
\epsscale{0.6}
\plotone{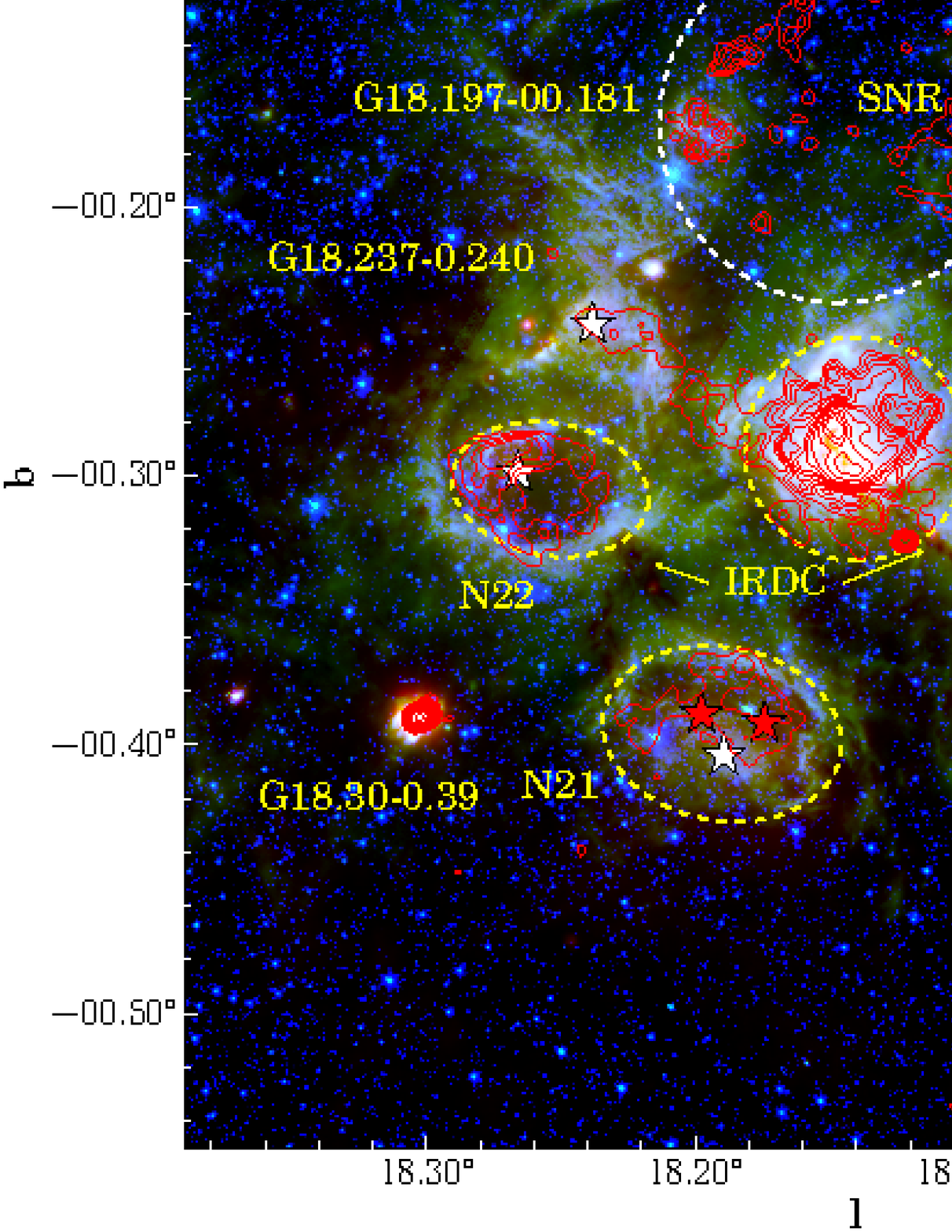}
\epsscale{0.6}
\plotone{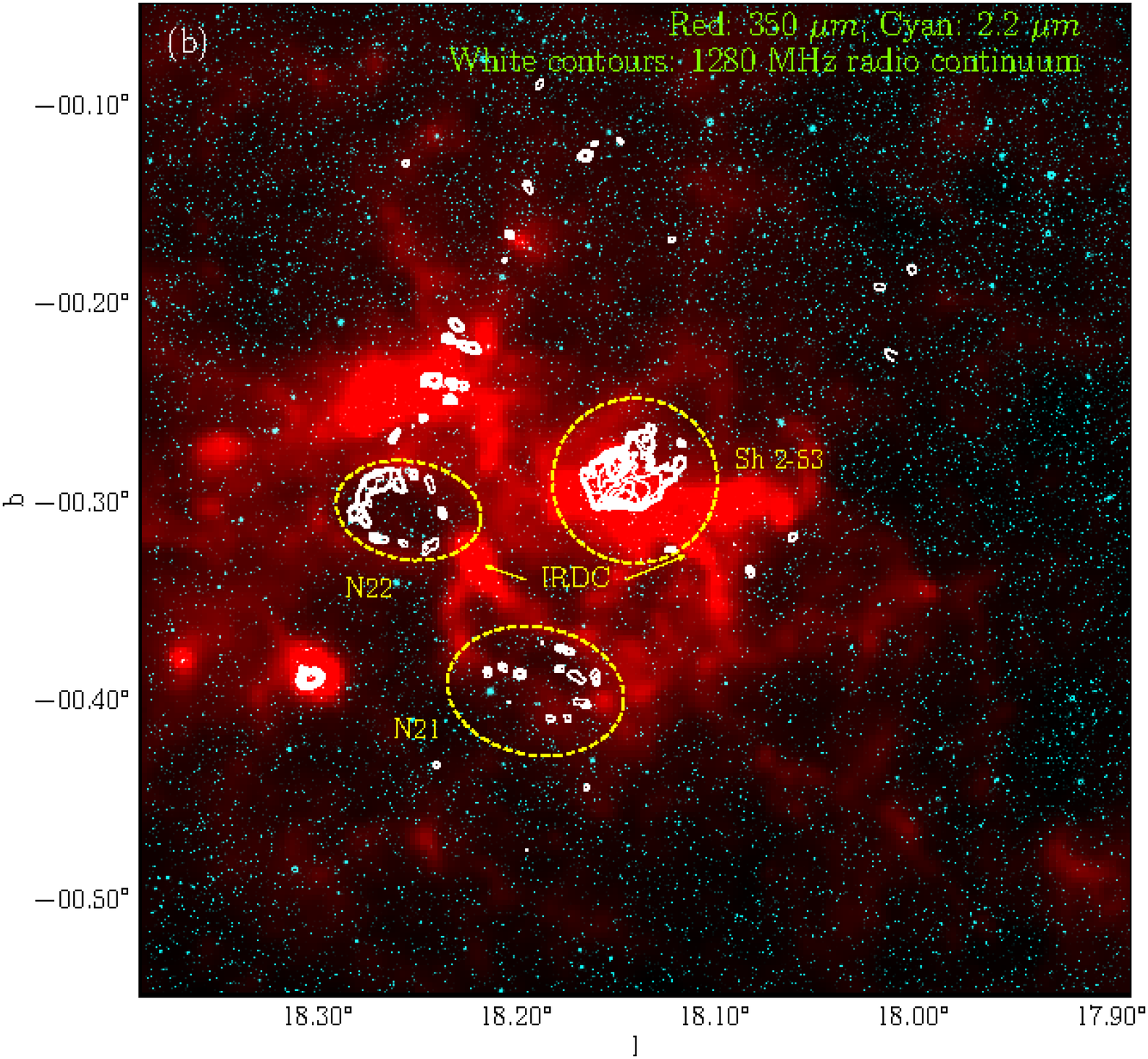}
\caption{\scriptsize (a) A color-composite image (red: 70 $\mu m$; green: 8.0 $\mu m$; blue: 3.6 $\mu m$) of 30$' \times$30$'$ area around the
 Sh 2-53 region. The 610 MHz GMRT radio contours at levels of 3, 4, 5, 6, 8, 10, 12, 15, 20, 25, 30, and 40 mJy are also overlaid on the image.
 Several \hII regions, two MIR Galactic bubbles, IRDCs and a SNR are seen along the line of sight. The positions of the bubbles, Sh 2-53 region,
 and SNR are marked by ellipses and circles, respectively. The positions of spectroscopically confirmed O-stars \citep{paron13} are also marked
 by white stars, while two more photometrically identified O-stars towards the N21 bubbles \citep{watson08} are marked by
 red stars. (b) A two color-composite image of the region (red: 350 $\mu m$; cyan: 2.2 $\mu m$). The 1280 MHz GMRT contours at levels of
 2.0, 2.5, 3.0, 5.0, 8.0, 12.0, 15.0, 20.0, 30.0, and 100.0 Jy are also overlaid on the image.}
\label{fig1}
\end{figure*}

A color-composite map of the complex (red: 70 $\mu m$; green: 8.0 $\mu m$; blue: 3.6 $\mu m$) of our selected 30$'$ $\times$ 30$'$
 area is presented in Figure~\ref{fig1}a. The 610 MHz radio continuum contours are also overlaid on the map. The selected area contains
 several important sources, i.e., two MIR bubbles \citep[N21 and N22;][]{churchwell06}, at least six \hII regions (see radio continuum
 contours towards G18.197-00.181, G18.237-0.240, G18.30-0.39, Sh 2-53, and the bubbles N21 and N22 in Figure~\ref{fig1}a), 
 infrared dark clouds (IRDCs), and the supernova remnant SNR G18.1-0.1 \citep{green14}. However, based on the local
 standard rest velocity (V$_{LSR}$), we find that not all these sources are physically linked with the S53 complex. Table~\ref{table1}
 lists the designations of these sources along with their Galactic coordinates, V$_{LSR}$, kinematic distances, and comments on their physical
 association with the S53 complex. \citet{paron13} identified three O-type stars towards this region which are marked by
 white stars in Figure~\ref{fig1}a. In this paper, we have adopted an average distance of 4.0 kpc for the S53 complex. In the
 following paragraphs, we have provided a brief description for only those sources which are associated with the S53 complex
 (see Table~\ref{table1}).

The MIR Galactic bubble N21 (l =18$\degr$.190, b =$-$0$\degr$.396) has been classified as a broken or incomplete ring with an average
 angular radius and thickness of 2$\farcm$16 and 0$\farcm$5 \citep{churchwell06}, respectively (also see Figure 1a in this paper). The
 distance to the bubble is reported to be 3.6 kpc \citep{anderson09}. The velocity of the radio recombination line associated with
 the bubble N21 is 43.2 km s$^{-1}$ \citep{lockman89}. One of the sources toward this bubble is spectroscopically identified to be a late
 O-type giant \citep{watson08}. Later, \citet{paron13} also confirmed this source as an O-type star using the optical spectrum.
 Two more sources that are photometrically identified as probable O-type stars \citep{watson08}, are also marked by red stars
 in Figure~\ref{fig1}a.

The bubble N22 (l = 18$\degr$.254, b =$-$0$\degr$.305) has a complete or closed ring-like appearance with an average angular radius of
 1$\farcm$69 \citep[which corresponds to $\sim$2.5 pc at a distance of 4.0 kpc;][]{churchwell06,anderson09}. It also encloses an \hII
 region. The velocity of the radio recombination line associated with the bubble N22 is 50.9 km s$^{-1}$ \citep{lockman89}.
 \citet{ji12} suggested a possibility of the interaction between the expanding \hII region linked with this bubble and the surrounding
 molecular clouds. Using the optical spectrum, \citet{paron13} identified an O-type star toward this bubble (see the white star
 in Figure~\ref{fig1}a).

The ionized gas linked with the Sh 2-53 region is traced in a velocity range of 50-53.9 km s$^{-1}$, and the distance to the region
 is reported to be 4.3 kpc \citep{blitz82,kassim89,kolpak03}. The velocity and the corresponding estimated distance are also in good
 agreement with the values reported for the molecular cloud associated with the ultra-compact \hII region, U18.15-0.28 \citep{anderson09}.
 \citet{paron13} identified three B-type sources towards this region using the spectroscopic observations.
 
Another \hII region in our selected area, G18.237-0.240, has a V$_{LSR}$ of 47 km s$^{-1}$ \citep{kassim89}, which also hosts a
 spectroscopically confirmed O-type star \citep[][marked in Figure~\ref{fig1}a]{paron13}. A similar velocity of this region
 as of the MIR bubbles indicates that they could be part of the same molecular cloud.  
 
The complex nature of this region has made it very much intriguing. Understanding the ongoing physical processes in such a complex region
 requires a thorough multi-wavelength analysis. We present a two color-composite image (red: 350 $\mu m$; cyan: 2.2 $\mu m$) of the region
 in Figure~\ref{fig1}b. The GMRT 1280 MHz radio contours are also overlaid on the image. The emission at 350 $\mu m$, a good tracer
 of cold dust, is mainly seen surrounding the MIR bubbles and the Sh 2-53 region. The IRDCs, which appear dark at shorter wavelengths
 ($<$70 $\mu m$), are visible at longer wavelengths (see Figure~\ref{fig1}b).

\section{Physical conditions towards the region}
\label{sec:surrclouds}
In this section, we first present the analysis of the radio continuum data, followed by the analysis of {\it Herschel} images to
 construct the column density and temperature maps for identification of cold condensations and the distribution of cold dust.
 The identification and clustering analysis of YSOs are presented in the final sub-section.
 
\subsection{Radio continuum emission and the dynamical age}
\label{sec:radio}
The strong Lyman continuum flux from massive stars ionizes the surrounding matter and develops \hII regions. The integrated radio
 continuum flux of an \hII region is often used to determine the spectral type of the ionizing source. We first estimate the radio continuum
 flux density at 1280 MHz and the size of the \hII region (see radio contours in Figure~\ref{fig1}) using the {\sc tvstat} task of
 AIPS. The corresponding Lyman continuum flux (photons s$^{-1}$) needed to develop each individual \hII region is estimated using
 the following equation from \citet{moran83}:
\begin{equation}
  S_{Lyc} = 8 \times 10^{43}\left(\frac{S_\nu}{mJy}\right)\left(\frac{T_e}{10^4 K}\right)^{-0.45}\left(\frac{D}{kpc}\right)^2 \left(\frac{\nu}{GHz}\right)^{0.1}
\label{lyman_flux}
\end{equation}

where the different parameters in the equations are followings. The symbol $\nu$ is the frequency of observations, $S_\nu$ is the integrated
 flux density, $T_e$ is the electron temperature, and $D$ corresponds to the distance to the source. In this calculation,
 it is assumed that all the \hII regions are homogeneous and spherically symmetric. They are also assumed to be classical
 \hII regions ($T_e$ of $\sim$10000 K), and a single source is responsible to develop each of them. The 1280 MHz integrated
 flux, corresponding Lyman continuum flux, and probable spectral types of the sources responsible to develop \hII regions associated
 with both the bubbles and the Sh 2-53 region are listed in Table~\ref{table2}. The estimated Lyman continuum flux for the
 bubble N21 is 10$^{48.23}$ photons sec$^{-1}$, which fits to a single ionizing source having spectral type of O8V for solar
 metallicity \citep{smith02}. This spectral type is also in agreement with the spectral type determined by \citet{paron13}.
 
Similarly, the estimated Lyman continuum fluxes for the other two \hII regions associated with the bubble N22 and Sh 2-53 are
 10$^{48.32}$ and 10$^{48.79}$ photons sec$^{-1}$, respectively, and these fluxes correspond to the ionizing sources
 of O8V and O7V stars, respectively. However, it should be mentioned here that the spectral types determined using the radio
 continuum flux are prone to large error depending on the adopted distance, and the size of the \hII region. Hence, the
 determinations of the spectral types using spectroscopic observations are always more robust. \citet{paron13} reported
 presence of an O- and a B-type stars towards the bubble N22, and three B-type sources towards the Sh 2-53 region.

The dynamical age of an \hII region helps us to understand if an \hII region is old enough to influence the
 formation of YSOs seen around it. The ionization front of an \hII region expands until an equilibrium is achieved between
 the rate of ionization and recombination. A theoretical extent of the \hII region, known as Str\"omgren radius
 \citep[$R_S$;][]{stromgren39}, can be computed with the following formula:
\begin{equation}
R_S = \left(\frac{3S_{Lyc}}{4\pi n_0^2\beta_2}\right)^{1/3}
\end{equation}
where $n_0$ is the initial ambient density, and $\beta_2$ is the recombination coefficient. The value of $\beta_2$ is taken
 to be 2.60$\times$10$^{-13}$ cm$^3$ s$^{-1}$ for a classical \hII region \citep{stahler05}. 
 
Once the ionized region developed, a shock front is generated at the interface of the ionized gas and the surrounding cold material
 because of the large difference in temperature and pressure. The shock front is further evolved with time by propagating into the
 surroundings. The radius of such an ionized region at any given time can be formulated as \citep{spitzer78}:
\begin{equation}
    R(t) = R_S \left(1 + \frac{7c_{II} t_{dyn}}{4R_S}\right)^{4/7}
\end{equation}

where $t_{dyn}$ is the dynamical age of the \hII region and c$_{II}$ is the sound speed in an \hII region which is
 11$\times$10$^5$ cm s$^{-1}$ \citep{stahler05}. We have estimated the dynamical ages of all three \hII regions towards the bubbles N21
 and N22, and the Sh 2-53 region. The radii of the \hII regions, R(t), were estimated by using the {\sc tvstat} task of the AIPS. The
 calculated dynamical age might vary substantially depending on the initial ambient density. Hence, the Str\"{o}mgren radius, and the
 dynamical age of the \hII regions are calculated for a range of ambient density from 1000 to 10000 cm$^{-3}$ \citep[i.e., from classical to
 ultra-compact \hII regions;][]{kurtz02}. The calculated dynamical ages for ambient densities of 1000, 2000, 5000, and 10000
 cm$^{-3}$ are also listed in Table~\ref{table2}. As can be seen in Table~\ref{table2}, dynamical ages for all the ambient densities
 are less than 1 Myr. However, it must be noted that the region is assumed to be uniform and spherically symmetric. Hence, the calculated
 dynamical ages of all the \hII regions should be treated as a qualitative value.
\subsection{Distribution of molecular gas and cold dust}
\label{sec:herschel}
\begin{figure*}
\epsscale{0.8}
\plotone{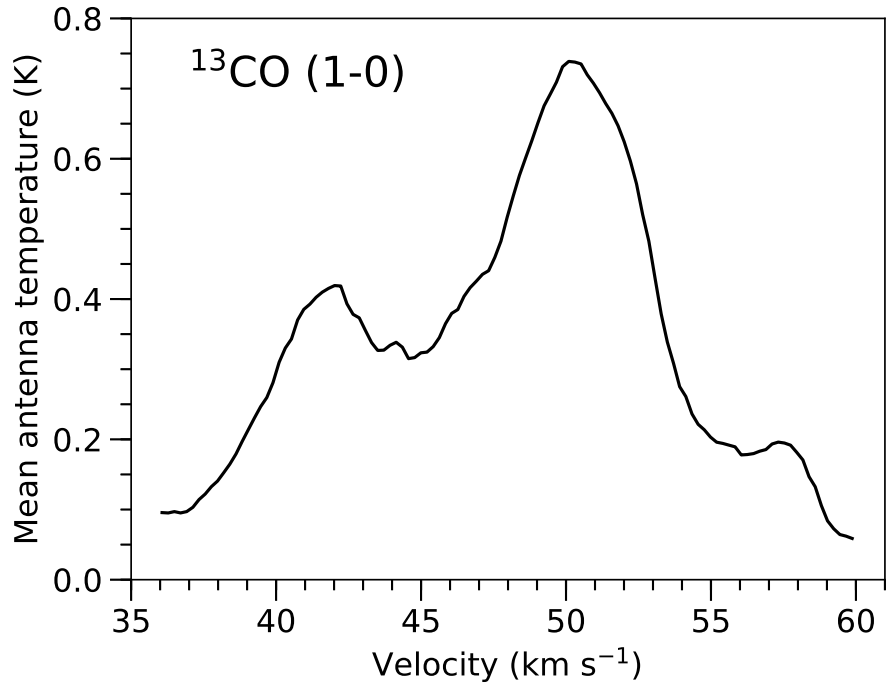}
\caption{The $^{13}$CO spectrum towards the S53 complex obtained by averaging the whole emission for the selected 30$'\times$30$'$ area
 centered at $l=$ 18$^\circ$.140, $b=$-0$^\circ$.300. Three peaks are clearly depicted in the spectrum.}
\label{fig2}
\end{figure*}

\begin{figure*}
\epsscale{1.0}
\plotone{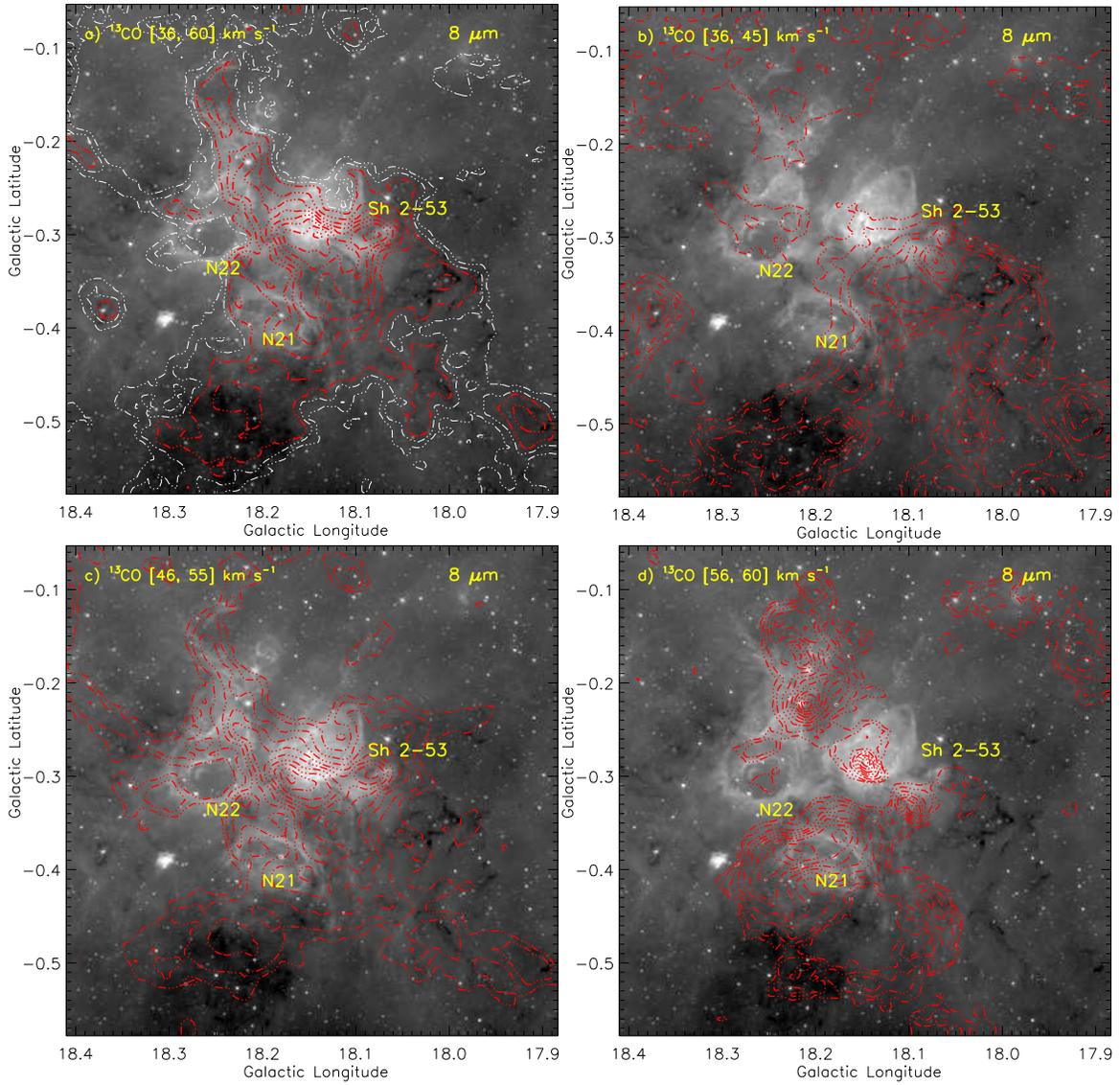}
\caption{(a) The velocity integrated $^{13}$CO contours for a velocity range from 36--60 km s$^{-1}$ are overlaid on the {\it Spitzer} 8
 $\mu m$ image. In the direction of the S53 complex, three molecular clouds are traced within this velocity range. (b)--(d) The velocity
 integrated contours for velocity ranges from 36--45, 46--55, and 56--60 km s$^{-1}$, respectively, overlaid on the {\it Spitzer} 8 $\mu m$
 image.}
\label{fig3}
\end{figure*}

A careful examination of the $^{13}$CO (J = 1--0) spectrum in the direction of the S53 complex reveals the presence of at least three
 molecular components in a velocity range of 36--60 km s$^{-1}$. Figure~\ref{fig2} shows the $^{13}$CO spectrum toward the S53
 complex. The spectrum is generated by averaging the whole emission from our selected field containing the S53 complex.
 As can be seen in the spectrum, three molecular cloud components are traced in the full velocity ranges of 36--45 km s$^{-1}$
 (hereafter MC18.20--0.50), 46--55 km s$^{-1}$ (hereafter MC18.15--0.28), and 56--60 km s$^{-1}$ (hereafter MC18.20--0.40). In
 Figure~\ref{fig3}a, the molecular emission integrated over the velocity range of 36--60 km s$^{-1}$ is overlaid on the {\it Spitzer}
 8 $\mu$m image. The $^{13}$CO emission contours integrated over three different velocity ranges (i.e. 36--45, 46--55, and 56--60 km
 s$^{-1}$) are also overlaid on the Spitzer 8 $\mu$m image, revealing the three molecular components
 (see Figures~\ref{fig3}b,~\ref{fig3}c, and~\ref{fig3}d)
 in the direction of our selected target field. It is to be noticed in Figure~\ref{fig3} that the Sh 2-53 region and the bubble
 N22 are mainly associated with MC18.20-0.50 and MC18.15-0.28 molecular clouds. However, the molecular gas associated with the
 bubble N21 is traced in the cloud MC18.20--0.40,  which is having a velocity range from 56--60 km s$^{-1}$. This velocity of
 molecular gas is higher than the velocity of the radio recombination line (V$_{LSR}$ $\sim$ 43.2 km s$^{-1}$) associated with the
 bubble N21 \citep{lockman89}. The discrepancy of about 10 km s$^{-1}$ between the velocity of the radio recombination line
 towards the bubble N21 and the velocity of the associated molecular gas can be explained by non-circular motion \citep{jones13}.
 Several molecular condensations are seen in the integrated intensity map (see in velocity integrated $^{13}$CO map of the region
 in Figure~\ref{fig3}a).

In order to examine the cold condensations toward the selected region, we have employed the {\it Herschel} images to construct the column density
 and temperature maps. We have performed a pixel-by-pixel modified blackbody fit on the {\it Herschel} 160, 250, 350 and 500 $\mu m$ images.
 The 70 $\mu m$ image is excluded as the flux in this band includes emission from the warm dust.
All the higher resolution images were convolved to the lowest resolution of 37$''$ (beam size of the 500 $\mu m$ image) after
 converting them to the same flux unit (i.e., Jy pixel$^{-1}$). The background flux was estimated from a nearby dark patch of the sky
 ($l=$ 18$^\circ$.50, $b=$ 0$^\circ$.86, for an area of 15$' \times$15$'$), and was subtracted from the corresponding image
 \citep[see][for more details]{mallick15}.
 
Finally, a modified blackbody was fitted on pixel-by-pixel of all four images by using the formula \citep{launhardt13}:
\begin{equation}
S_\nu (\nu) - I_{bg} (\nu) = B_\nu(\nu, T_d) \Omega (1 - e^{-\tau(\nu)})
\end{equation}
where optical depth can be written as:
\begin{equation}
\tau(\nu) = \mu_{H_2} m_{H} \kappa_\nu N(H_2)
\end{equation}
where the notations are as follows: $S_\nu (\nu)$ is the observed flux density, $I_{bg}$ corresponds to the background flux density,
 $B_\nu(\nu, T_d)$ is the Planck's function, $T_d$ stands for the dust temperature, $\Omega$ is the solid angle subtended by a pixel, $\mu_{H_2}$
 represents the mean molecular weight, $m_H$ stands for the hydrogen mass, $\kappa_\nu$ is the dust absorption coefficient, and $N(H_2)$ is the
 column density. Here, we assumed a gas-to-dust ratio of 100 and used the following values in the calculation: $\Omega$ = 4.612$\times$10$^{-9}$
 steradian (i.e. the area of a pixel of 14$\arcsec\times$14$\arcsec$), $\mu_{H_2}$\,=\,2.8 and $\kappa_\nu$\,=\,0.1~$(\nu/1000~{\rm GHz})^{\beta}$
 cm$^{2}$ g$^{-1}$, and a dust spectral index ($\beta$) of 2 assuming sources are having thermal emission in the optically thick medium
 \citep{hildebrand83}.

\begin{figure*} 
 \epsscale{0.6}
\plotone{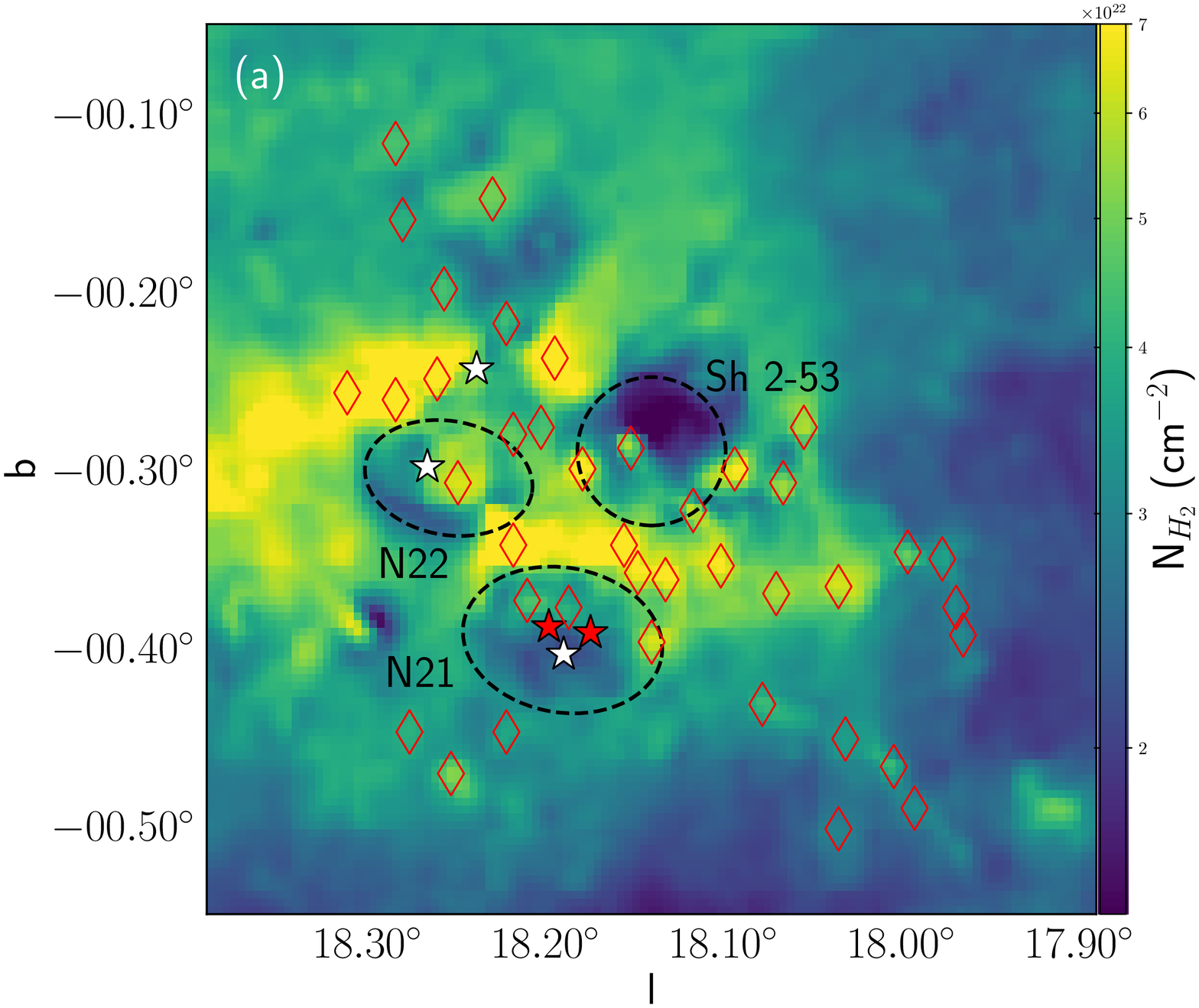}
 \epsscale{0.6}
\plotone{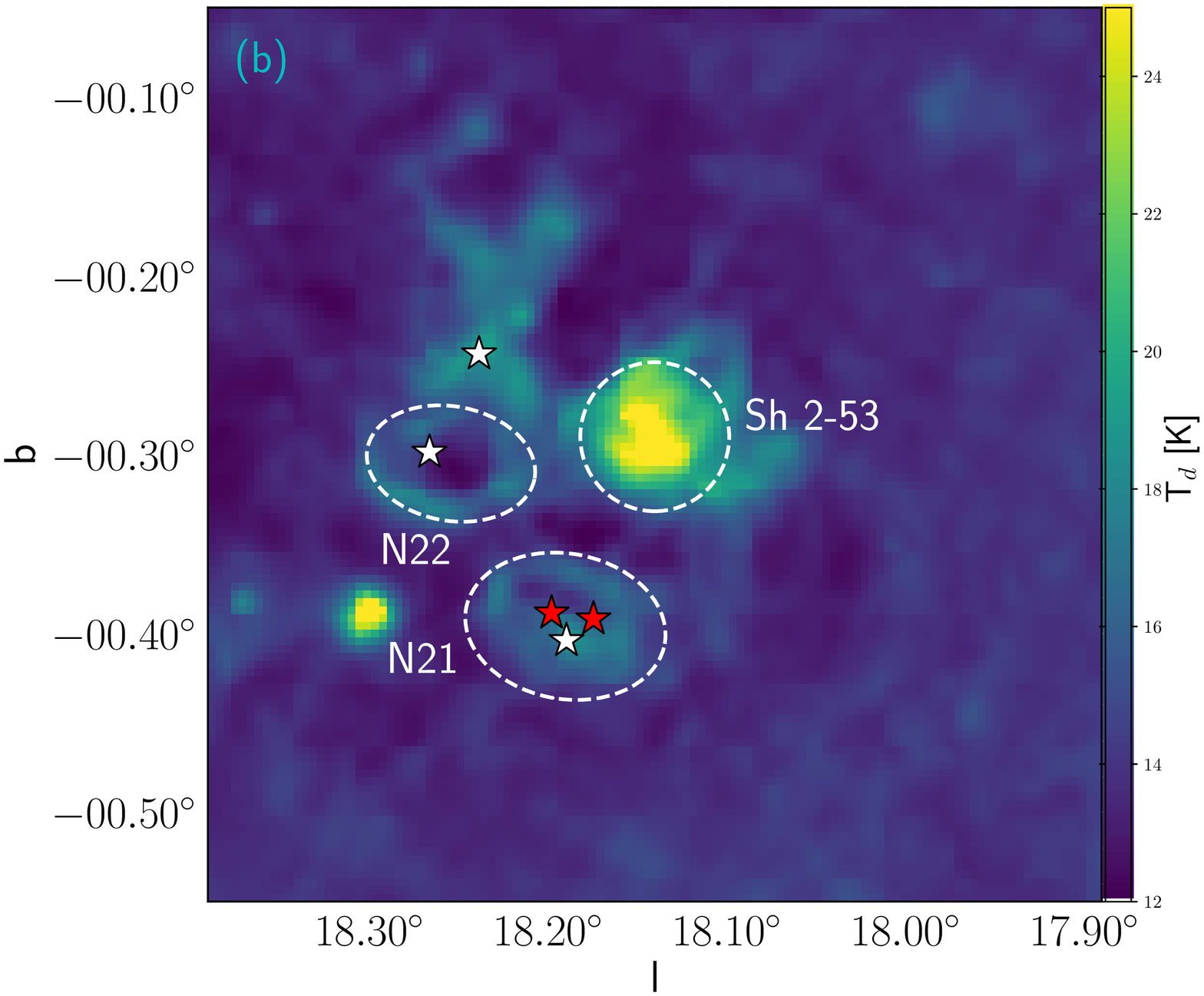}
\caption{\scriptsize {\it Herschel} (a) column density and (b) temperature maps of the S53 complex. Several
 cold clumps are also marked ($\diamond$) in the column density map. The remaining symbols are similar to those in Figure~\ref{fig1}. }
\label{fig4}
\end{figure*}

The final column density and temperature maps of the region are presented in Figures~\ref{fig4}a and~\ref{fig4}b, respectively. Identification of
 the molecular clumps and estimation of their column densities are performed using the {\sc clumpfind} software \citep{williams94}. The mass of each
 clump is estimated using the formula \citep{mallick15}:
\begin{equation}
M_{clump} = \mu_{H_2} m_H Area_{pix} \Sigma N(H_2)
\end{equation}
where $Area_{pix}$ is the area subtended by a single pixel.

A total of 72 clumps are identified toward the 30$' \times$30$'$ area of the region. However, a careful visual examination of the
 spatial distribution of these identified clumps with respect to the integrated $^{13}$CO map reveals that only 40 clumps are associated
 with MC18.20-0.50, MC18.15-0.28 and MC18.20-0.40 clouds (see Figure~\ref{fig4}a where all 40 clumps are marked by diamond symbols).
 The masses of the identified {\it Herschel} clumps range from 3.0$\times$10$^3$ -- 3.0$\times$10$^4$ M$_\odot$. These calculated
 masses of {\it Herschel} clumps are generally high, and possibly be over estimated because of the presence of multiple molecular
 clouds at different velocities along the line-of-sight (all are not discussed in this paper) which are integrated in the
 {\it Herschel} column density map. One of the massive condensations with a mass of about 2.3$\times$10$^4$ M$_\odot$ is identified
 toward the junction of the bubbles, which is reported to be a collapsing massive prestellar core \citep[see clump \#5 of ][]{zhang17}.
\subsection{Young stellar population}
It is always required to have a detailed knowledge of YSOs and their clustering behavior in a given star-forming region to characterize the
 areas of the ongoing star formation. Hence, we have identified YSOs in our selected area of the S53 complex using the NIR and MIR
 color-magnitude and color-color schemes. Furthermore, we have performed the nearest neighbor analysis to look for the clusterings of YSOs, and
 hence, the areas of active star formation. An elaborative description of the adopted schemes to identify YSOs is given below.
\subsubsection{Selection of YSOs}
We have utilized four different MIR/NIR schemes to identify and classify YSOs among the point sources detected in the selected
 30$'\times$30$'$ area. Note that these schemes are not mutually exclusive. In fact, the successive scheme(s) may include
 YSOs that are identified in the previous scheme(s). We have arranged the schemes as per their robustness to classify the YSOs.
 If the same source is identified in multiple schemes with different classes then preference is given to the class characterized
 in the preceding scheme.

\begin{figure*}
\begin{tabular}{cc}
    \includegraphics[width=0.45\textwidth]{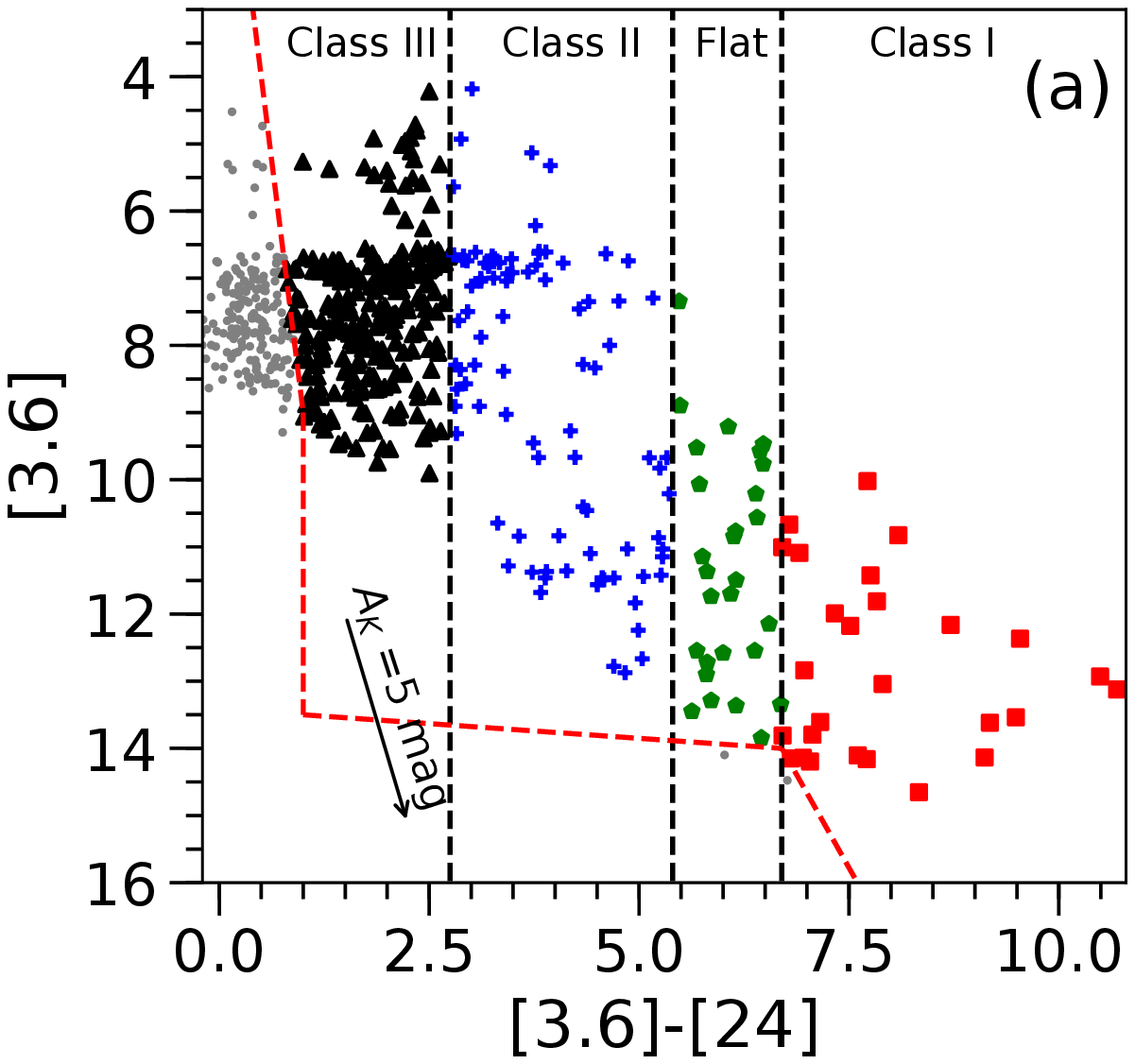} & \includegraphics[width=0.45\textwidth]{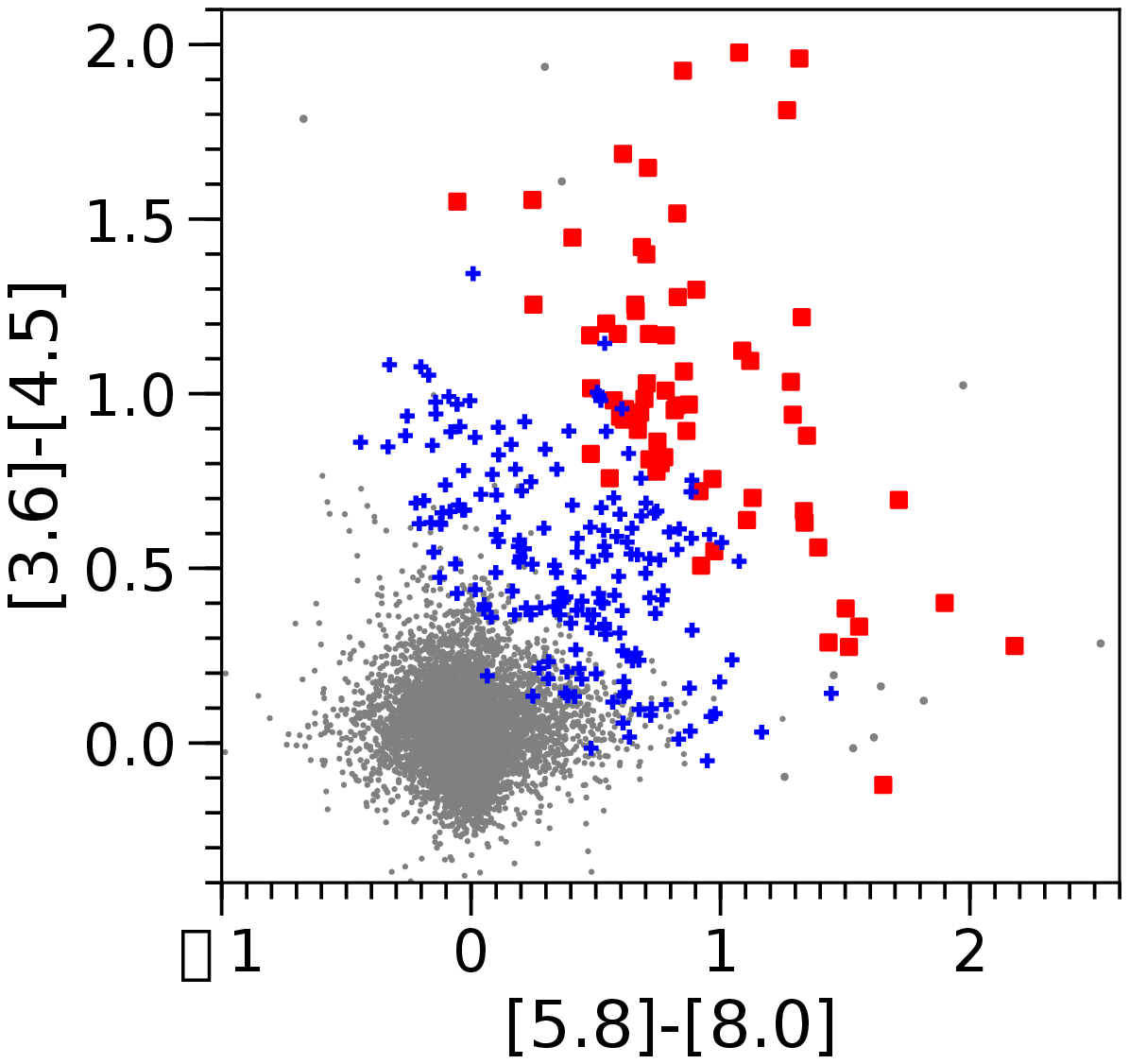} \\ \hfill
    \includegraphics[width=0.45\textwidth]{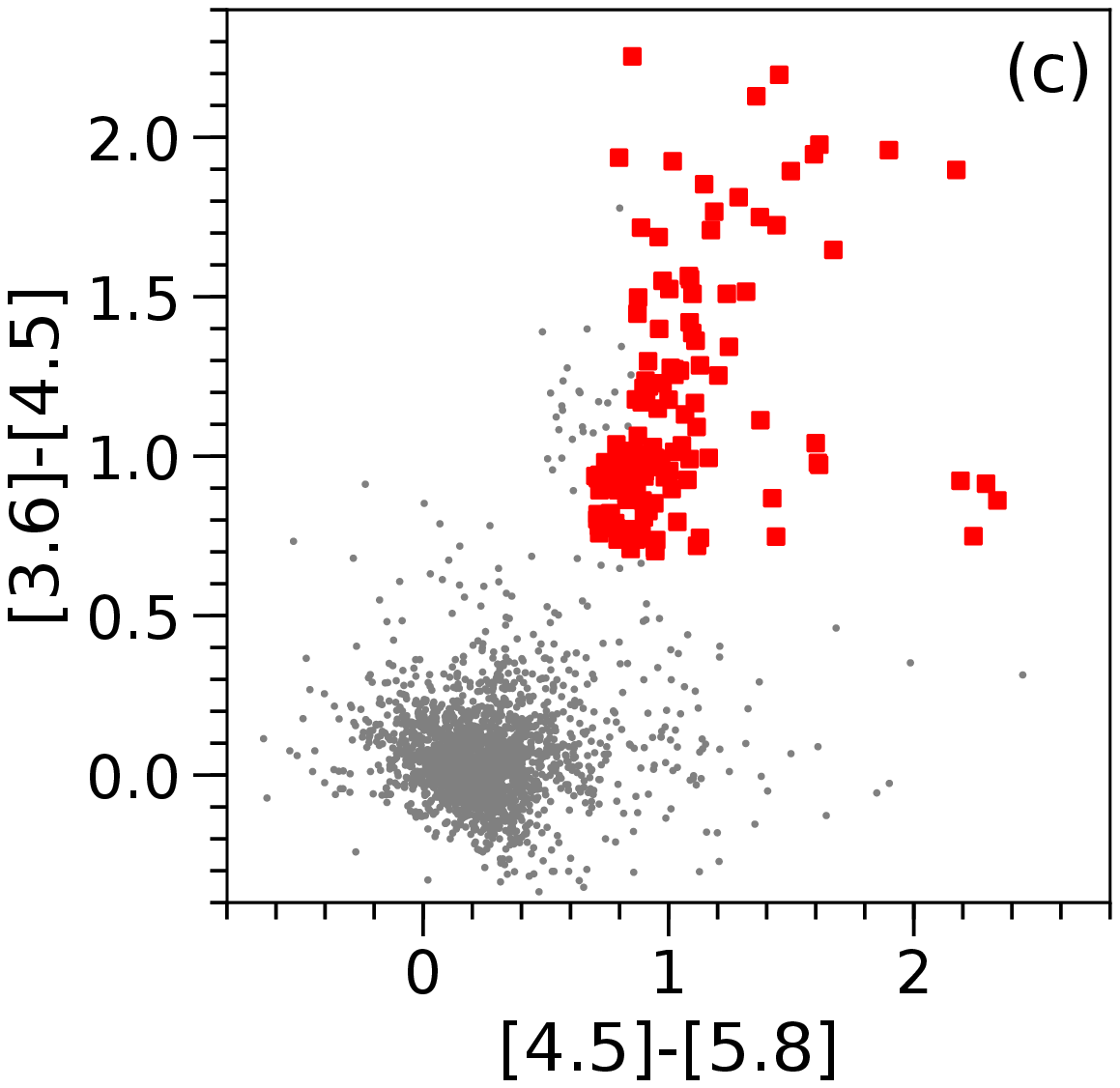} & \includegraphics[width=0.45\textwidth]{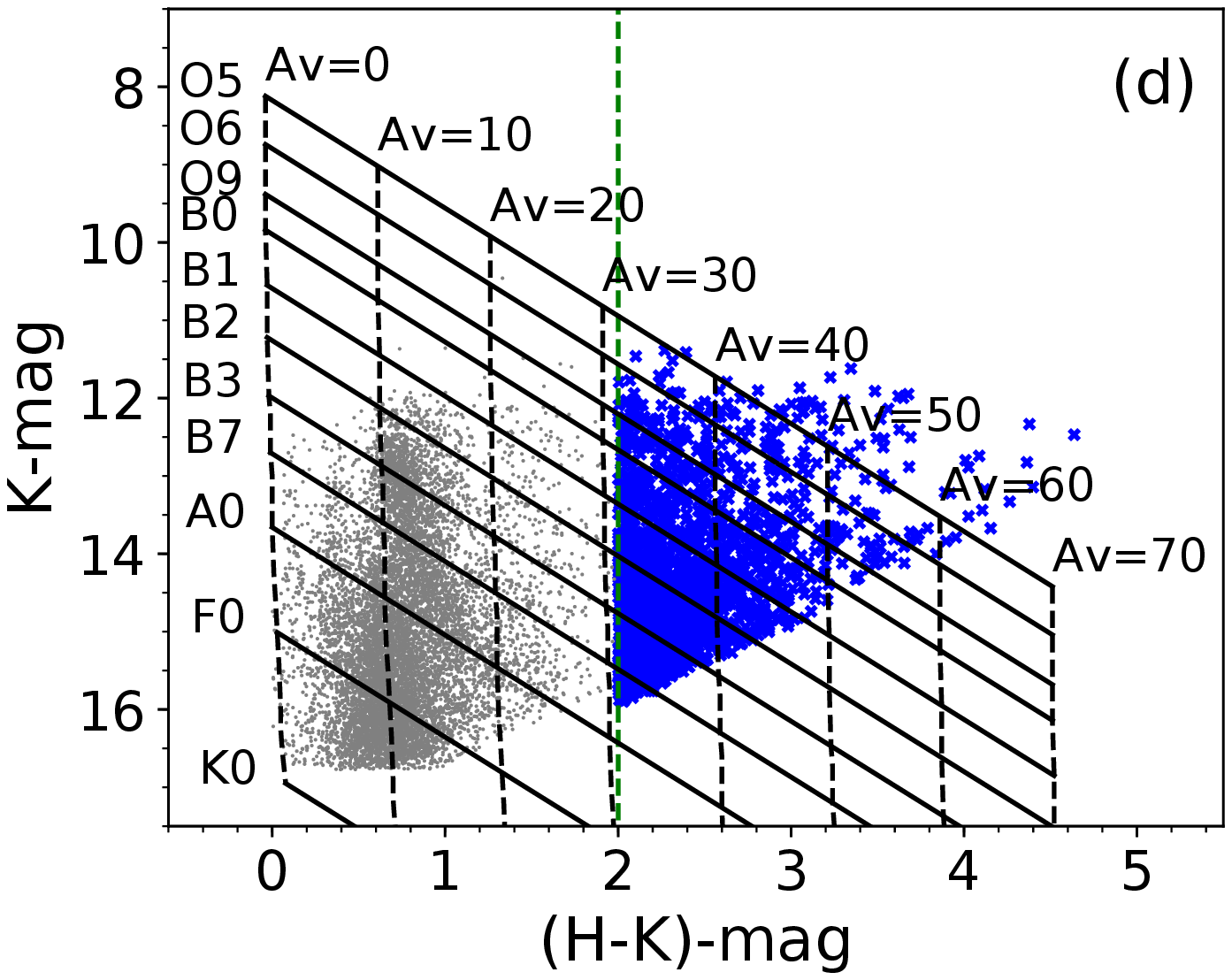} \\
\end{tabular}
\caption{\scriptsize The selection schemes of YSOs using the MIR and NIR color-color and color-magnitude criteria. (a) The [3.6]$-$[24]/[3.6]
 color-magnitude diagram of the point-like sources towards the region. Class I sources are marked by filled red squares. Flat-spectrum, Class II
 and Class III sources are marked by green filled pentagons, blue crosses, and black filled triangles, respectively. An extinction vector
 corresponding to $A_K$ of 5 mag is also shown. The sources in this scheme are selected and classified into different classes following
 the criteria outlined in \citet{guieu10} and \citet{rebull11}. The red lines separates the YSOs from the field stars. (b) The color-color
 ([5.8]$-$[8.0]/[3.6]$-$[4.5]) diagram of point-like
 sources detected in all four {\it Spitzer}-IRAC bands. Elimination of possible contaminants and classification of remaining point sources
 are performed based on the scheme described in \citet{gutermuth09}. (c) The color-color ([4.5]$-$[5.8]/[3.6]$-$[4.5]) diagram of those sources
 that are detected only in the first three {\it Spitzer}-IRAC bands among the four bands. The sources are classified following the criteria
 given in \citet{hartmann05} and \citet{getman07}. (d) The NIR color-magnitude (H$-$K/K) diagram of all the reliable NIR sources detected
 in the $H$ and $K$ bands (see text for more details). A cutoff value of H$-$K of 2.0 mag was obtained from a nearby reference field
 ($12\arcmin \times 12\arcmin$ area centered at $l$ = 18$^\circ$.12, $b$ = 0$^\circ$.04).}
\label{fig5}
\end{figure*}

1. We first employed MIR color-magnitude scheme to separate out young sources. We found a total of 570 sources to be common in 3.6 and
 24 $\mu$m bands. The [3.6]$-$[24]/[3.6] color-magnitude diagram of these 570 point sources is shown in Figure~\ref{fig5}a. The color
 criteria given in \citet{guieu10} and \citet{rebull11} were adopted to separate out different classes of YSOs. Finally, we identified
 a total of 144 YSOs (27 Class I, 28 Flat-spectrum and 89 Class II), and 246 Class III sources following this particular scheme.

2. The MIPSGAL 24 $\mu m$ images of star-forming regions are often suffered from strong nebulosity, and more number of point
 sources are expected to detect in the {\it Spitzer}-IRAC bands compared to 24 $\mu m$ images. Therefore, we have also
 constructed a [5.8]$-$[8.0])/([3.6]$-$[4.5] color-color diagram of the point sources to identify YSOs (see Figure~\ref{fig5}b).
 The YSOs identified using the criteria given in \citet{gutermuth09}, are categorized into different evolutionary stages based
 on their slopes of the spectral energy distribution (SED) in the IRAC bands \citep[i.e. $\alpha_{IRAC}$; see][for more details]{lada06}.
 Accordingly, a total of 249 YSOs (71 Class I and 178 Class II), and 4158 Class III sources were identified using this scheme.

3. Nebulosity also affects the IRAC 8.0 $\mu m$ band, and hence, many point sources may not be seen in the 8.0 $\mu m$ image. Therefore,
 in addition to the schemes mentioned above, the [3.6]$-$[4.5]/[4.5]$-$[5.8] color-color diagram was also constructed to identify
 YSOs (see Figure~\ref{fig5}c). All the sources with [4.5]$-$[5.8] $\geq$ 0.7 mag and [3.6]$-$[4.5] $\geq$ 0.7 mag in this color-color
 diagram are considered as Class I YSOs \citep{hartmann05,getman07}. A total of 140 Class I YSOs were identified following this scheme.

4. In general, YSOs appear to be much redder than the nearby field stars in the NIR color-magnitude diagram because of the presence of
 circumstellar material. Hence, we have also constructed a NIR color-magnitude diagram (H$-$K/K) of the point sources to
 identify YSOs (Figure~\ref{fig5}d). It is assumed in this scheme that all the sources above a certain $H-K$ color cut-off are probable
 YSOs. The $H-K$ color cut-off of 2.0 was estimated from the H$-$K/K color-magnitude diagram of a nearby field region ($l=$18$^\circ$.12,
 $b=$0$^\circ$.04; FoV: 12$'\times$12$'$). Using this scheme, a total of 3023 red sources were identified above the cut-off value.

\begin{figure*} 
 \epsscale{0.6}
\plotone{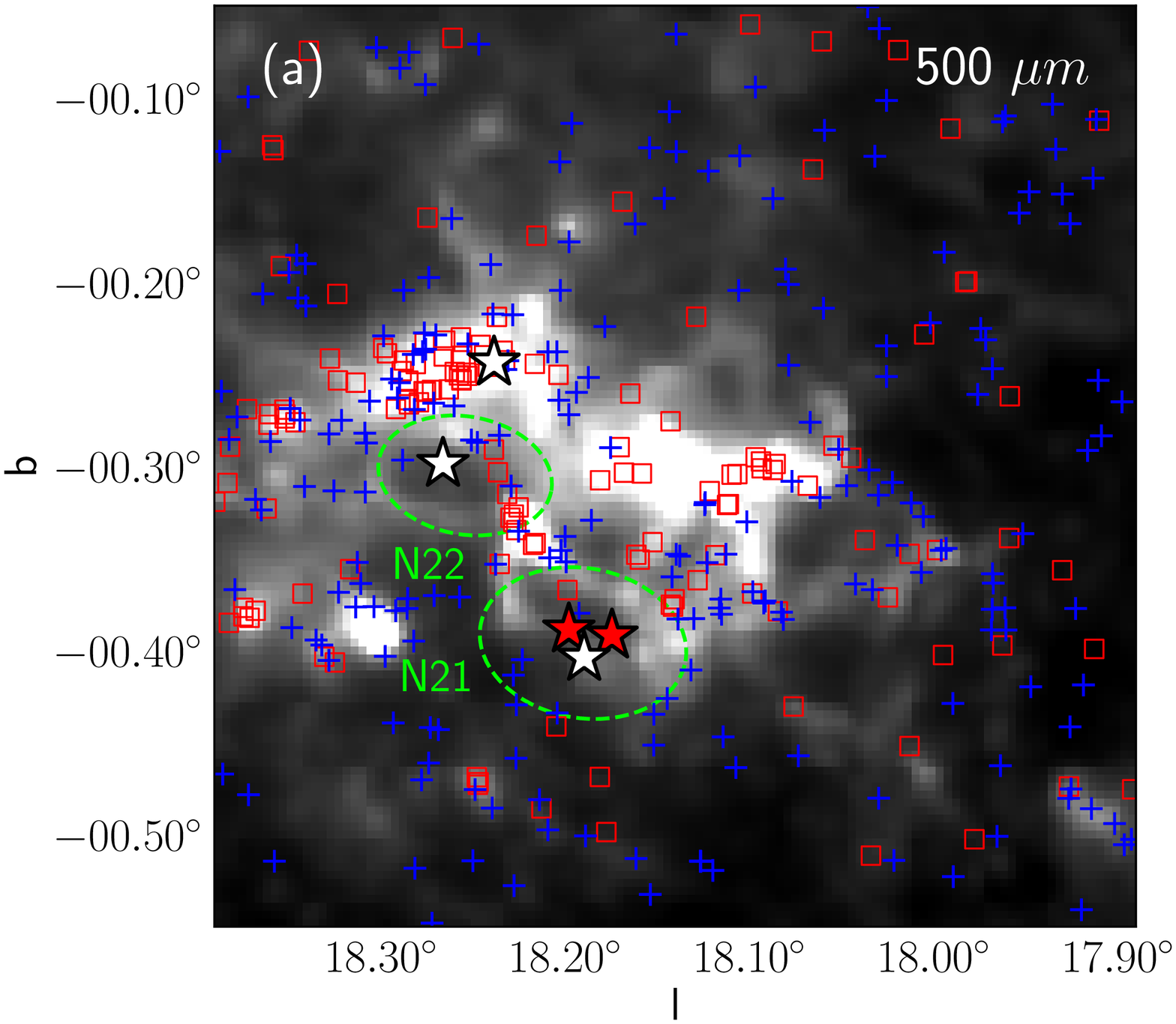}
 \epsscale{0.6}
\plotone{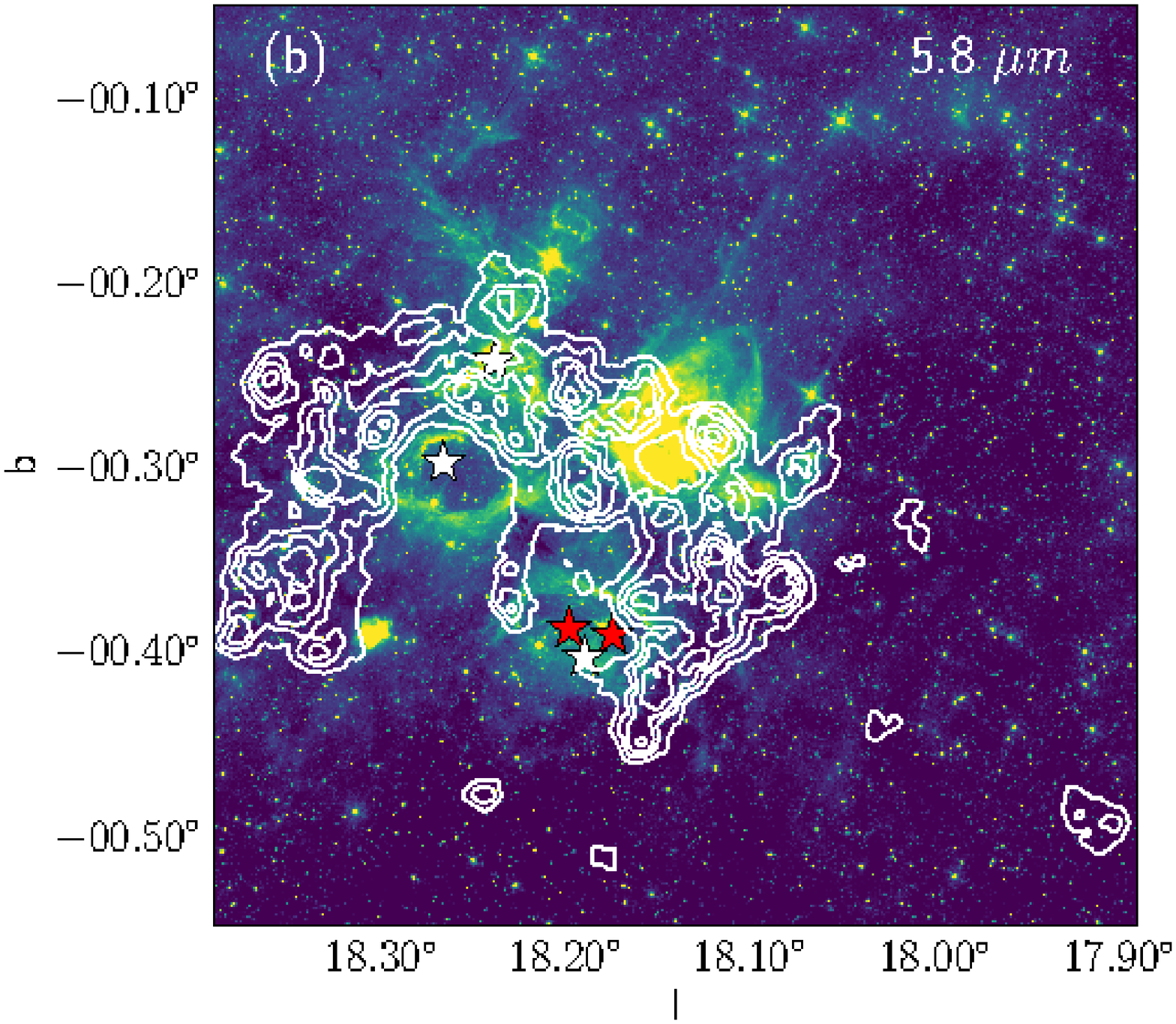}
\caption{\scriptsize (a) Class I and Class II YSOs, marked by red squares and blue crosses, respectively, overlaid on the 500 $\mu m$
 image. The YSOs identified using the NIR scheme are not shown here, however, they are used in the surface density analysis.
 (b) Overlay of 20NN surface density contours (white) of all the YSOs on the {\it Spitzer}-IRAC 5.8 $\mu m$ image. The surface
 density contours of YSOs are drawn at 5, 7, 9, 12, 16, 20, 25 and 30 YSO pc$^{-2}$. YSOs are mainly clustered towards the junction
 of the Sh 2-53 region and the bubbles (see text for more details).}
 \label{fig6} 
\end{figure*}

As mentioned before that it is not necessary that YSOs identified using four different schemes are mutually exclusive. Hence, to have
 a complete catalog of YSOs, they were cross-matched. Accordingly, we have found a total of 393 YSOs (i.e., 139 Class I, 28 Flat spectrum,
 and 226 Class II), 4260 Class III sources, and 2691 red sources (identified using NIR color-magnitude scheme) in our selected
 30$'\times$30$'$ region. All the identified Class I and Class II YSOs are marked on the 500 $\mu m$ image of the region
 (see Figure~\ref{fig6}a). It can be seen in the figure that YSOs, mainly Class I sources, are situated towards the periphery
 of the bubbles, and the Sh 2-53 region.

\subsubsection{Surface density analysis of YSOs}
\label{sec:surface_density}
Clustering of YSOs in a given region helps us to identify the areas of active star formation. To examine the clustering behavior
 of YSOs toward the S53 complex, we have performed a 20-nearest-neighbor (20NN) surface density analysis of identified YSOs as
 it is shown by \citet{schmeja08} that 20NN surface density analysis is capable enough to identify clusters of 10--1500 YSOs.
 Note that in this analysis, all the YSOs are assumed to be located at the same distance of 4.0 kpc. The surface density contours
 drawn at 5, 7, 9, 12, 16, 20, 25, and 30 YSOs pc$^{-2}$, overlaid on the 5.8 $\mu m$ image are shown in Figure~\ref{fig6}b. It
 can be seen in the figure that YSOs are mainly clustered around the \hII regions (i.e., N21, N22 and Sh 2-53 regions). A cluster
 of YSOs is also seen toward the Galactic east of the bubble N22. However, this particular cluster of YSOs is associated with the
 molecular cloud in the velocity range of 66--76 km s$^{-1}$, and hence, they have no physical association with
 the S53 complex having a velocity range of 36--60 km s$^{-1}$.
 
\section{Molecular filaments and Star formation activity}
\label{sec:SF}
Earlier studies \citep{paron13, watson08, ji12} reported that the ionizing feedback from massive stars has influenced the star
 formation activity toward the bubbles N21 and N22, and the Sh 2-53 region. But the estimated dynamical ages of these \hII regions
 (see Table~\ref{table2}) are not consistent enough to influence the formation of Class I and Class II YSOs.
 For example, with a typical ambient density of 1000 cm$^{-3}$, the calculated dynamical ages of the \hII regions
 ($\lesssim$ 0.3 Myr) suggest that they might not have induced the formation of Class I and Class II YSOs having average ages of 0.46 and 1-3 Myr,
 respectively \citep{evans09}. However, it must be noted that the derived dynamical ages of the \hII regions are qualitative
 values. Hence, a possibility of the influence of the ionized gas on the star formation in surrounding molecular cloud cannot be ruled out
 totally. Yet, we have analyzed the $^{13}$CO data in detail to examine the physical condition and the dynamics of the molecular
 gas, which might be helpful to understand other possible mechanisms for the ongoing star formation activity.

\begin{figure*}
\epsscale{0.6}
\plotone{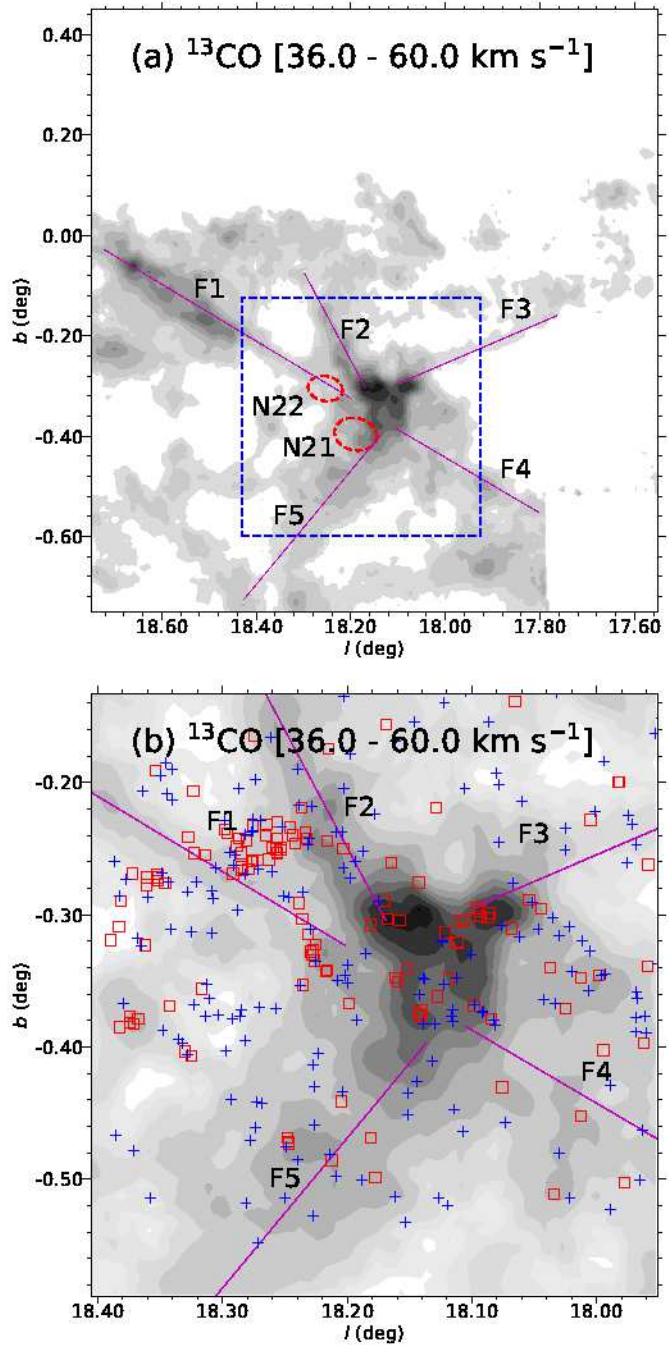}
\caption{\scriptsize (a) The velocity integrated $^{13}$CO map of a large 1$^\circ$.2$\times$1$^\circ$.2 area around the S53 complex for a
 velocity range of 36--60 km s$^{-1}$. Five filaments from F1--F5 are identified in this integrated molecular map, marked by magenta
 lines. Positions of the bubbles N21 and N22 are also marked and labeled. (b) The zoomed in view of $^{13}$CO map of the S53 complex,
 marked by a blue dashed box in panel (a). The positions of the Class I (red squares) and Class II (blue crosses) YSOs are also marked.}
\label{fig7}
\end{figure*}

\begin{figure*}
\epsscale{0.45}
\plotone{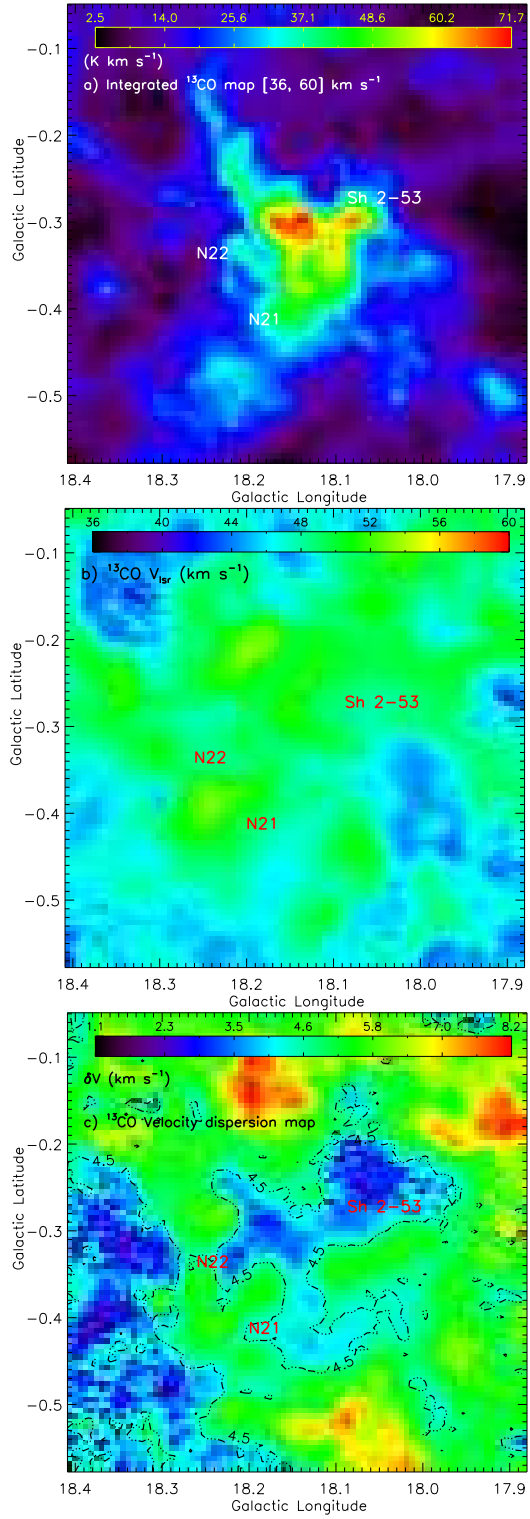}
\caption{\scriptsize (a) The velocity integrated $^{13}$CO map of the central region around the S53 complex for the velocity range of
 36--60 km s$^{-1}$. (b) First moment or the velocity map of the region. (c) Second moment map or the velocity dispersion map of the 
 S53 complex. Dispersion in the gas velocity can be clearly noted towards the S53 complex.}
\label{fig8}
\end{figure*}

The velocity integrated $^{13}$CO intensity map for a larger 1$^\circ$.2$\times$1$^\circ$.2 area for a velocity range from
 36--60 km s$^{-1}$, is presented in Figure~\ref{fig7}a. A zoomed in view of intensity map for the central part of the complex is
 also presented in Figure~\ref{fig7}b. Presence of at least five molecular filamentary structures having typical length of five to
 ten parsecs are identified in the velocity integrated $^{13}$CO map (see marked filaments labeled as F1-F5 in Figures~\ref{fig7}a,b).
 The filaments are found to be connected to the central ``hub'' hosting the S53 complex (see Figure~\ref{fig7}b). The positions
 of Class I and Class II YSOs are also marked in Figure~\ref{fig7}b. Though YSOs are primarily seen towards the central ``hub'',
 few YSOs are also found towards the filamentary structures (see F1, F2, and F5 for example). We have constructed the moment maps of the central
 part of the complex which are presented in Figure~\ref{fig8}. The molecular zeroth, first, and second moment maps are shown in
 Figures~\ref{fig8}a, \ref{fig8}b, and \ref{fig8}c, respectively. The zeroth moment map (or integrated intensity map) is similar to
 that shown in Figure~\ref{fig7}b. The first-order moment map is the measure of the intensity-weighted mean velocity of the emitting
 gas (see Figure~\ref{fig8}b). The velocity dispersion map is usually represented by the second-order moment map. A large velocity
 dispersion is seen towards the S53 complex. Generally, a large dispersion could arise due to a broad single velocity component and/or
 may indicate the presence of two or more narrow components with different velocities along the line of sight (see Figure~\ref{fig8}c).
 In Figure~\ref{fig8}c, we have also over-plotted a velocity dispersion contour with a level of 4.5 km s$^{-1}$. This map clearly
 indicates a high dispersion (i.e. 2--4.5 km s$^{-1}$) towards the S53 complex.
 
\begin{figure*}
\epsscale{0.5}
\plotone{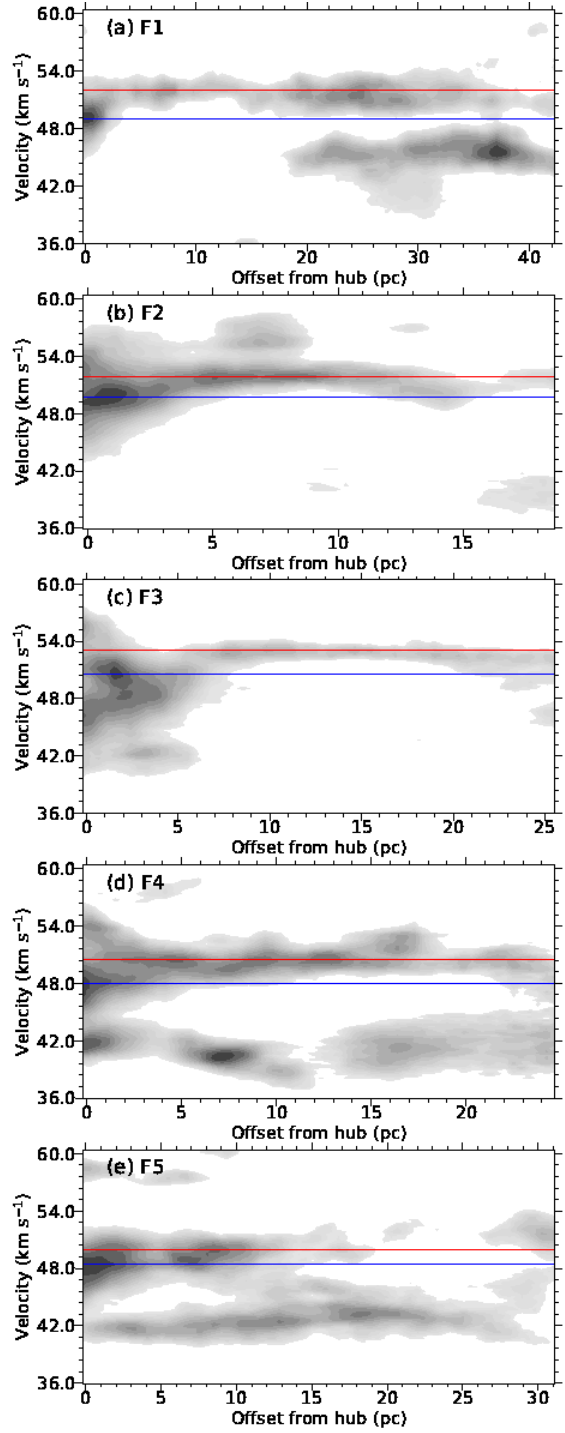}
\caption{\scriptsize (a)-(e) The \pv diagrams for filaments F1--F5 marked in Figure~\ref{fig7}. The peak velocities corresponding
 to the filaments and the central ``hub'' are marked by red and blue lines, respectively. Gradient in velocities in a range from
 2-3 km s$^{-1}$ is noted almost for all the filaments within 5 pc towards the central ``hub''. The broadening of the velocity
 profile can also be easily noted in the \pv diagrams as these filaments move to the central ``hub''.}
\label{fig9}
\end{figure*}

We have also constructed \pv diagrams for all five identified filaments separately for the velocity range from 36--60 km s$^{-1}$
 (see Figure~\ref{fig9}). All the \pv diagrams are constructed along the filaments, and the distance of each filament is measured
 in parsecs from the central ``hub'', which hosts the S53 complex. Note that a single heliocentric distance of 4.0 kpc is assumed for all five
 filaments to calculate their lengths. Also, in this analysis, no projection effects or inclination angles have been taken into
 account, and all of them are assumed to be projected on the plane-of-sky. The peak velocities corresponding to each filament and
 the central ``hub'' are marked by red and blue lines, respectively.

As the S53 complex is located at the junction of at least five molecular filaments (see Figures~\ref{fig7}a,b), hence, there is a strong
 possibility that the star formation in the S53 complex is influenced by these filaments. We, therefore, search for the possibility of
 active role of these filaments in the formation of the S53 complex. A gradient in the peak velocity at a range from 2-3 km s$^{-1}$ can
 be easily noticed almost in all the filaments, within 5 pc length near the central ``hub'' except for F5 (see marked lines in all the
 subplots of Figure~\ref{fig9}). Note the identified velocity gradients are significantly higher than the velocity resolution of GRS
 (0.21 km s$^{-1}$). Velocity gradients seen in all the filaments are also an order higher than the sound speed of $\sim$0.2 km s$^{-1}$
 at a typical filament temperature of about 10 K \citep{omukai05}. Such supersonic velocity gradients imply that there could be significant
 substructures along the filaments \citep[see][ for more details]{andre16, hacar13}.
 
Gradients in the velocity could also be interpreted by rotation. However, the presence of a substantial velocity gradients
 in all the filaments which are oriented at different directions from the central ``hub'', discard any such possibility. 
 Such ``hub--filament'' configurations are also seen in the simulations of a magnetized cloud. The dissipation of the
 magnetized cloud causes the cloud to condense along its field lines into a dense layer \citep[see][]{nakamura08}. It has
 been also discussed in \citet{dobashi92} that the filaments may form because of the twisting of the magnetized gas. In
 this case, the filamentary cloud rotates along its major axis that gives rise the gradient in the velocity of the
 filaments towards the central ``hub''. The broadening of the velocity profiles of all the filaments is also noted as they
 move towards the central ``hub''. It has been discussed by \citet{olmi16,nakamura14,kirk13} that filaments feeding a central
 ``hub'' generally show a velocity gradient towards the central ``hub'' and might appear with a larger velocity dispersion
 towards the ``hub''. This result indicates that the molecular cloud (velocity range of 36--60 km s$^{-1}$), hosting the
 S53 complex, is possibly fed by at least five filaments. Also, an active star formation process is noted towards this
 central ``hub'' by the presence of cold clumps and YSO clusters (see Sections~\ref{sec:herschel} and \ref{sec:surface_density}).
 Though not many YSOs are seen along the filaments, YSOs are, however, found to be clustered towards the central ``hub''
 (see Figure~\ref{fig7}b). Several such ``hub-filament'' systems are reported in the literature which also occasionally host
 massive stars \citep[see e.g.,][]{baug15,dewangan15, peretto13,schneider12}.

Overall, from the observational signatures, it seems that the star formation activity in the S53 complex is possibly influenced by
 channeling of matter along five molecular filaments towards the S53 complex.
\section{Conclusions}
\label{sec:conclusions}
In this work, we have carried out a detailed multi-wavelength analysis of a selected 30$'$ $\times$ 30$'$ area around the Sh 2-53 region
 (i.e., S53 complex) to probe the ongoing star formation activities. Several authors reported that the star formation activity
 towards this region is influenced by the ionizing feedback of massive star. However, we find a different mechanism to be operating in this
 S53 complex than those reported in the literature. The major findings of the study are the following.

1. The molecular cloud hosting the Sh 2-53 region is well traced in a velocity range of 36--60 km s$^{-1}$ which also harbors
 two MIR bubbles (namely, N21 and N22) and three \hII regions. At least three molecular components are identified having
 velocity ranges of 36--45, 46--55, and 56--60 km s$^{-1}$ that host the S53 complex.
 
2. Considering our estimates of Lyman continuum photons using the GMRT 1280 MHz data, we find that the primary ionizing sources for the \hII
 regions linked with the Sh 2-53 region and both the bubbles are O7V and O8V stars, respectively. This result is also in agreement with the spectral
 type of the ionizing sources reported in the literature.

3. The dynamical ages of the \hII regions associated with the bubbles and the Sh 2-53 region are estimated to be $\lesssim$0.3 Myr for a
 typical ambient density of 1000 cm$^{-3}$. Hence, it appears that they might not be capable enough for triggering the formation of Class I
 and Class II YSOs on their surrounding cloud.

4. Using the $^{13}$CO line data, at least five molecular filaments are identified in the integrated intensity map, and they appear to be
 radially directed to the central molecular condensation. It resembles more of a ``hub-filament" system, and the central molecular
 condensation contains the S53 complex.

5. The {\it Herschel} column density map traces several condensations in the selected area around the S53 complex. A massive
 clump (M$_{clump}$ $\sim$2.3 $\times$ 10$^{4}$ M$_{\odot}$) is identified toward the intermediate area between the bubbles N21 and N22.
 YSOs identified using the NIR and MIR photometric schemes, are also found to be clustered toward this central molecular
 condensation. It is suggestive of active star formation in this ``hub-filament" system.

6. The identified molecular filaments have noticeable velocity gradients (i.e. 2--3 km s$^{-1}$) as they move towards the central ``hub''.
 Such supersonic velocity gradients indicate the presence of significant substructures along the filaments. All these filaments also
 show a wider velocity profile towards the ``hub". These are indicative of molecular gas flow towards the central ``hub" along these filaments.
 
Based on our observational findings, we conclude that molecular filaments have influenced the formation of the massive stars
 and clusters of YSOs in the S53 complex, by channeling the molecular gas to the central ``hub'' hosting the complex.
\acknowledgments
We thank the anonymous referee for the constructive comments which have helped to improve the scientific content and the
 presentation of the paper. We thank the staff of IAO, Hanle and CREST, Hosakote, who made these observations possible. The facilities
 at IAO and CREST are operated by the Indian Institute of Astrophysics, Bangalore. We thank the staff of the GMRT that made these
 observations possible. GMRT is run by the National Centre for Radio Astrophysics of the Tata Institute of Fundamental Research. This
 work is based on data obtained as part of the UKIRT Infrared Deep Sky Survey. This publication made use of data products from the Two
 Micron All Sky Survey (a joint project of the University of Massachusetts and the Infrared Processing and Analysis Center/ California
 Institute of Technology, funded by NASA and NSF), archival data obtained with the {\it Spitzer} Space Telescope (operated by the Jet
 Propulsion Laboratory, California Institute of Technology under a contract with NASA). This publication makes use of molecular line
 data from the Boston University-FCRAO Galactic Ring Survey (GRS). The GRS is a joint project of Boston University and Five College
 Radio Astronomy Observatory, funded by the National Science Foundation (NSF) under grants AST-9800334, AST-0098562, and AST-0100793.
 LKD is supported by the Department of Space, Government of India. MT is supported by the JSPS KAKENHI Grant Number 15H02063. 
%

\begin{deluxetable*}{lccccc}
\tablewidth{0pt}
\tabletypesize{\scriptsize} 
\setlength{\tabcolsep}{0.070in}
\tablecaption{The details of different sources and their association with the S53 complex.\label{table1}}
\tablehead{
\colhead{Name of}    &\colhead{l}     &\colhead{b}     & \colhead{$V_{LSR}$}     &\colhead{Kinematic}     &\colhead{Physical Association}  \\  
\colhead{the source} &\colhead{(deg)} &\colhead{(deg)} & \colhead{(km s$^{-1}$)} &\colhead{Distance (kpc)}&\colhead{with the S53 complex}}
\startdata
Bubble N21                         &   18.190       & --0.396        &    43.2   &    3.6\tablenotemark{a}   & Yes    \\
Bubble N22                         &   18.254       & --0.305        &    50.9   &    4.0\tablenotemark{a}   & Yes    \\
Sh 2--53                           &   18.210       & --0.320        &    52.0   &    4.3\tablenotemark{b}   & Yes    \\
G18.237--0.240                     &   18.237       & --0.240        &    47.0   &    --                     & Yes    \\
SNR G18.1--0.1\tablenotemark{c,d,e}&   18.106       & --0.195        &   100.0   &    5.6\tablenotemark{d,e} & No     \\
G18.197--0.181\tablenotemark{f}    &   18.197       & --0.181        &     --    &   12.4\tablenotemark{f}   & No     \\
G18.30--0.39\tablenotemark{b}      &   18.303       & --0.390        &    32.3   &    2.8\tablenotemark{g}   & No           
\enddata
\tablenotetext{a}{\citet{churchwell06}} \tablenotetext{b}{\citet{kolpak03}} \tablenotetext{c}{\citet{green14}} \tablenotetext{d}{\citet{leahy14}}
\tablenotetext{e}{\citet{kilpatrick16}}
\tablenotetext{f}{\citet{anderson09}}
\tablenotetext{g}{\citet{wienen12}}
\end{deluxetable*}

\begin{deluxetable*}{lccccccc}
\tablewidth{0pt}
\tabletypesize{\scriptsize} 
\setlength{\tabcolsep}{0.065in}
\tablecaption{The dynamical ages of the \hII regions associated with the S53 complex seen in the 1280 MHz radio continuum map.
\label{table2}}
\tablehead{
\colhead{Name of the}&\colhead{Integrated 1280 MHz} &\colhead{Log of Lyman continuum}&\colhead{Spectral}&\multicolumn{4}{c}{Dynamical Age (Myr) for ambient density of} \\  
\colhead{\hII region}&\colhead{Flux (Jy)}           &\colhead{Flux (photons s$^{-1}$)}&\colhead{type}   &\colhead{1000 cm$^{-3}$}      &\colhead{2000 cm$^{-3}$}&\colhead{5000 cm$^{-3}$}&\colhead{10000 cm$^{-3}$} }
\startdata
N21                  &            1.30              & 48.23                    & O8V & 0.29 & 0.43 & 0.69 & 0.99 \\
N22                  &            1.60              & 48.32                    & O8V & 0.28 & 0.41 & 0.66 & 0.94 \\
Sh 2-53              &            4.74              & 48.79                    & O7V & 0.18 & 0.27 & 0.45 & 0.65 
\enddata
\end{deluxetable*}

\end{document}